\documentclass[12pt]{article}

\usepackage{amsmath,amsfonts,amssymb}

\def\tE{\tilde{E}}

\newcommand{\bD}{\mathbf{D}}

\def\bA{\mathbf{A}}

\def\bB{\mathbf{B}}

\def\tx{\tilde{x}}

\def\be{\begin{equation}}
\def\bD{\mathbf{D}}
\def\halpha{\hat{\alpha}}
\def\hbeta{\hat{\beta}}
\def\ee{\end{equation}}
\def\bD{\mathbf{D}}
\def\bea{\begin{eqnarray}}

\def\eea{\end{eqnarray}}

\def\tr{\mathrm{tr}\, }

\def\tr{\mathrm{Tr}}

\def\str{\mathrm{Str}}

\newcommand{\mU}{\mathcal{U}}

\def \bA{\mathbf{A}}

\newcommand{\tphi}{\tilde{\phi}}

\begin{document}

	\begin{titlepage}

		\vskip 0.4 cm
		
		\begin{center}
			{\Large{ \bf
					
				Remark About T-duality of Dp-Branes}}
			
			\vspace{1em}  Josef Kluso\v{n}$\,^1$
			\footnote{Email address:
				klu@physics.muni.cz}\\
			\vspace{1em} $^1$\textit{Department of Theoretical Physics and
				Astrophysics, Faculty
				of Science,\\
				Masaryk University, Kotl\'a\v{r}sk\'a 2, 611 37, Brno, Czech Republic}\\

			%
			%
			
			\vskip 0.8cm
			
		\end{center}

		\begin{abstract}
			This note is devoted to the analysis of T-duality of
			Dp-brane when we perform T-duality along directions that
			are transverse to world-volume of Dp-brane.
			
		\end{abstract}
		
		\bigskip
		
	\end{titlepage}
	
	\newpage

\section{Introduction and Summary}\label{first}
It is well known that T-duality is central property of string theory, 
for review, see for example \cite{Giveon:1994fu}.  Generally, if we consider
string sigma model in the background with metric $G_{MN}$ and NSNS two form $B_{MN}$ together
with dilation $\phi$ and this background possesses an isometry along $d-$directions we 
find that it is equivalent to string sigma model in T-dual background with dual fields $\tilde{G}_{MN},\tilde{B}_{MN}$ and $\tphi$ that are related to the original fields by famous Buscher's rules \cite{Buscher:1987sk,Buscher:1987qj} for generalization to more directions, see for example \cite{Ginsparg:1992af,Quevedo:1993vq}.

It is well known that string theories also contain another higher dimensional objects that transform non-trivially under T-duality. In this note we  focus on  Dp-branes \cite{Polchinski:1995mt,Polchinski:1996na}, for more recent review, see \cite{Simon:2011rw}. Originally Dp-brane was defined with the open string description where the string embedding coordinates obey $p+1$ Neumann boundary conditions and $9-(p+1)-$Dirrichlet ones at the boundary of the string world-sheet
\cite{Polchinski:1995mt}
\footnote{We consider Dp-brane in supersymmetric Type IIA or Type IIB theories where the critical dimension of target space time is $10$. Note that $p=0,2,4,6,8$ for Type IIA theory while $p=1,3,5,7,9$ for Type IIB theory.}. It was also shown by Polchinski that Dp-brane transforms into D(p+1)-brane when T-duality is performed along direction transverse to world-volume of Dp-brane and Dp-brane transforms to D(p-1)-brane in case when T-duality is performed along direction that Dp-brane wraps. In other words Dp-brane transforms with very specific way under T-duality transformations. 

On the other hand it is remarkable that many aspects of Dp-brane dynamics can be described by its low energy effective action which  is famous Dirac-Born-Infeld action \cite{Polchinski:1995mt}. Then one can ask the question whether this description of Dp-brane dynamics  could give correct description of T-duality transformation of Dp-brane. This situation is relatively straightforward when we perform T-duality along directions which Dp-brane wraps. This property is known as covariance of Dp-brane action under T-duality transformations as was previously studied in  
\cite{Kamimura:2000bi,Simon:1998az}. We generalize this approach to the T-duality along more longitudinal directions in the next section. 

It is important to stress that  in order to show  full covariance of Dp-brane action with respect to T-duality transformation we should also study how Dp-brane effective action changes when we perform T-duality along transverse directions to its world-volume. The goal of this paper is to perform such an analysis. Our approach is based on previous works that consider description of $N$ Dp-branes on the circle \cite{Taylor:1996ik}, for review see \cite{Taylor:1997dy}. It was shown there that such a configuration should be described by infinite number of Dp-branes on $\mathbf{R}$ which is covering space of $\mathbb{S}^1$ when we impose appropriate quotient conditions \cite{Taylor:1997dy}. Since this description was performed in the context of Matrix theory 
\cite{Banks:1996vh} the low energy effective action describing dynamics of $N-$ Dp-branes was Super Yang-Mills theory (SYM) defined on $p+1$ dimensional world-volume. Then it was shown in 
\cite{Taylor:1996ik} that this  theory transforms under T-duality into (SYM) defined on $p+2$ dimensional world-volume in T-dual background.  

The goal of this paper is to generalize this analysis to the case of  full DBI action for Dp-brane in the general background when we study T-duality along transverse directions. It is well known that such a T-duality can be defined when the target space-time fields do not depend on these coordinates explicitly. 
Since, following previous works, we should consider generalization of DBI action that describes infinite number of Dp-branes in covering space. 
Such an action is non-abelian generalization of DBI action that was introduced in    \cite{Myers:1999ps}. Then we follow very nice analysis performed in \cite{Brace:1998ku}. Explicitly, we introduce quotient conditions and solve them in the same way as in  \cite{Brace:1998ku}. We show that non-abelian action for infinite number of Dp-branes transforms to the action for D(p+d)-brane where $d-$ is number of T-dual directions in the T-dual background where T-dual background fields are related to the original one by generalized Buscher's rules \cite{Ginsparg:1992af,Quevedo:1993vq}.

Let us outline our results. We study how Dp-brane transforms under T-duality we consider T-duality either along longitudinal or transverse directions to Dp-brane' world-volume. We show that in the first case 
it transforms do D(p-d)-brane while in the second one it transforms to D(p+d)-brane when all background fields transform according to generalized Buscher's rules. This fact nicely shows covariance of Dp-brane under T-duality transformations. 

This paper is organized as follows. In the next section (\ref{second}) we introduce Dp-brane action and study T-duality along longitudinal directions. Then in section (\ref{third}) we consider T-duality along directions transverse to Dp-brane world-volume.

\section{Longitudinal T-duality}\label{second}
In this section we review T-duality transformation of Dp-brane when we perform T-duality
along $d-$ longitudinal directions. Explicitly, let us consider
DBI action in the general background with the 
 metric $G_{MN},B_{MN}$ and dilaton $\phi$. This
action has the form
\begin{equation}
S=-T_p \int d^{p+1}\xi e^{-\phi}
\sqrt{-\det (G_{\alpha\beta}+B_{\alpha\beta}+\lambda F_{\alpha\beta})}
\end{equation}
where
\begin{equation}
\lambda=2\pi\alpha' \ ,  \quad   T_p=\frac{2\pi}
{\lambda^{(p+1)/2}} \ . 
\end{equation}
where we also defined pull back of $G_{MN}$ and $B_{MN}$ defined as
\begin{equation}
	G_{\alpha\beta}=G_{MN}\partial_\alpha x^M\partial_\beta x^N \ , 
	\quad 
	 B_{\alpha\beta}=B_{MN}\partial_\alpha x^M
	 \partial_\beta x^N
\end{equation}
where $\xi^\alpha,\alpha=0,1,\dots,p$ label world-volume directions of Dp-brane and where $x^M,M=0,1,\dots,9$ parametrize embedding of DBI action in the target space-time. 

Now we would like to perform T-duality along last $d-$directions when we presume that there are directions which Dp-brane wraps. The fact that these directions are longitudinal mean that Dp-brane world-volume coordinates coincide with the target space ones. Explicitly we have
\begin{equation}
x^m=\xi^m \ , \quad m=9-d,\dots,9 \ . 
\end{equation}
Then we presume that all world-volume fields do not depend on $\xi^{\halpha}$ only where $\halpha=0,1,\dots,p-d$. Let us also introduce matrix $E_{MN}=G_{MN}+B_{MN}$. 
Then we have
\begin{eqnarray}
E_{\alpha\beta}+\lambda F_{\alpha\beta}=
\left(\begin{array}{cc} E_{\halpha\hbeta}+\lambda F_{\halpha\hbeta} &
E_{\halpha n}+\lambda \partial_{\halpha }A_n \\
E_{m\hbeta}-\lambda \partial_{\hbeta}A_m & E_{mn} \\ 
\end{array}\right) \ ,
\end{eqnarray}
where $E_{\halpha\hbeta}=E_{\mu\nu}\partial_{\halpha}x^\mu\partial_{\hbeta}x^\nu \ , \quad 
\mu,\nu=0,1,\dots,9-d$.
Then performing standard manipulation with determinant we obtain
\begin{eqnarray}
& &\det (E_{\alpha\beta}+\lambda F_{\alpha\beta})=\nonumber \\
& &\det \left(E_{\halpha\hbeta}+\lambda F_{\halpha\beta}-
(E_{\halpha m}+\lambda \partial_{\halpha}A_m)
\tE^{mn}(E_{n\hbeta}-\lambda \partial_{\hbeta}A_n)\right)
\det E_{mn}
\nonumber \\
& &=\det\left(E_{\halpha\hbeta}-E_{\halpha m}\tE^{mn}E_{n\hbeta}+\lambda F_{\halpha\hbeta}+
\right.
\nonumber \\
& & \left.\lambda E_{\halpha m}\tE^{mn}\partial_{\hbeta}A_n
-\lambda \partial_{\halpha} A_m \tE^{mn}E_{n\hbeta}+
\lambda^2 \partial_{\halpha}A_m \tE^{mn}\partial_{\hbeta}A_n\right)\det  E_{mn} \ , 
\nonumber \\
\end{eqnarray}
 and 
where $\tE^{mn}$ is inverse to $E_{mn}$ in the sense that
$\tE_{mn}E^{nk}=\delta_m^k$. As the next step we define  $T-$dual coordinates
\begin{equation}
\tx_m\equiv \lambda A_m \ .
\end{equation}
Then we can write
\begin{eqnarray}
& &E_{\halpha\hbeta}-E_{\halpha m}\tE^{mn}E_{n\hbeta}+\partial_{\halpha}x_m
\tE^{mn}\partial_{\hbeta}x^n=\nonumber \\
& &\partial_{\halpha}x^\mu (E_{\mu\nu}-E_{\mu m}\tE^{mn}E_{n\nu})\partial_{\hbeta}x^\nu+
\partial_{\halpha}\tx_m \tE^{mn}\partial_{\hbeta}\tx_n \  , \nonumber \\
& &E_{\halpha m}\tE^{mn}\partial_{\hbeta}\tx_n=\partial_{\halpha}x^\mu E_{\mu m}
\tE^{mn}\partial_{\hbeta}\tx_n \ , \nonumber \\
& &-\partial_{\halpha}\tx_m \tE^{mn}E_{n\hbeta}=
-\partial_{\halpha}\tx_m \tE^{mn}E_{n\nu}\partial_{\hbeta}x^\nu 
\nonumber \\
\end{eqnarray}
that can be interpreted as an embedding of D(p-d)-brane in T-dual background with the background fields 
\begin{eqnarray}\label{trTdualback}
& &\tE_{\mu\nu}=E_{\mu\nu}-E_{\mu m}E^{mn}E_{n\nu} \ , \nonumber \\
& &\tE_{\mu}^{ \ m}=E_{\mu n}E^{nm} \ , \quad  \tE^m_{ \ \nu}=
-E^{mn}E_{n\nu} \ ,  \nonumber \\
& &e^{-\tphi}=e^{-\phi}\det E_{mn} \ . \nonumber \\
\end{eqnarray}
Explicitly, the D(p-d)-brane action in T-dual background has the form 
\begin{equation}
S=-T_{p-d}\int_0^{\sqrt{\lambda}}d^{d}\xi
\int d^{p-d+1}\xi e^{-\tphi}\sqrt{-\det (\tE_{\halpha\hbeta}+\lambda F_{\halpha\hbeta})} \ , 
\end{equation}
where 
\begin{equation}\label{Tduallon}
\tE_{\halpha\hbeta}=\tE_{\mu\nu}\partial_{\halpha}x^\mu \partial_{\hbeta}x^\nu+
\tE_{\mu}^{ \ m}\partial_{\halpha}x^\mu \partial_{\hbeta}\tx_m
+\tE^m_{ \ \nu}\partial_{\halpha}\tx_m \partial_{\hbeta}\tx^\nu+
\partial_{\halpha}\tx_m \tE^{mn}\partial_{\hbeta}\tx^n \ , 
\end{equation}
and where we defined tension for D(p-d)-brane in the form 
\begin{equation}
T_{p-d}=T_p\int_0^{\sqrt{\lambda}} d^d\xi=
\lambda^{d/2} T_p \ . 
\end{equation}
Note that the transformation rules for T-dual fields given in 
(\ref{trTdualback}) coincide with the results derived previously 
in  \cite{Ginsparg:1992af,Quevedo:1993vq} and which are now derived
independently using covariance of Dp-brane under T-duality transformations.

However in order to see consistency of T-duality covariance of Dp-branes
we should also consider opposite situation when we consider Dp-brane in 
general background and perform T-duality along directions that are transverse
to the world-volume of Dp-brane.

\section{Transverse T-duality}\label{third}
In this section we consider opposite situation when we study Dp-brane
in the background that has isometry along $d-$directions in the transverse space to Dp-brane world-volume. The best way how to describe such a Dp-brane in to consider infinite number of Dp-branes on the covering space of torus $T^d$ which is $\mathbf{R}^d$ and impose appropriate quotient conditions. 
Further, we should also consider appropriate action for $N$ Dp-branes which is famous Mayer's non-abelian action  \cite{Myers:1999ps}
\begin{equation}\label{actnon}
S=-T_p\str\int d^{p+1}\xi e^{-\phi}
\sqrt{-\det (P[E_{\alpha\beta}]+P[E_{\alpha r}E^{rs}((Q^{-1})_s^t-\delta_s^t)E_{t\beta}]+
	\lambda F_{\alpha\beta})\det Q^i_j} \ , 
\end{equation}
where $i,j,k,l,m,n,r,s,t,\dots$$=p+1,\dots,9$ label directions transverse to the world-volume
of $N$ Dp-branes. Note that the location of $N-$ Dp-branes in the transverse space is determined by $N\times N$ Hermitean matrices $\Phi^m,m=p+1,\dots,9$ and all background fields depend on them so as for example $E_{\alpha\beta}(\Phi)$ and so on. We use convention where
$\Phi^m$ are Hermitean matrices and field strength $F_{\alpha\beta}$ is defined as
\begin{equation}
F_{\alpha\beta}=\partial_\alpha A_\beta-\partial_\beta A_\alpha+i[A_\alpha,A_\beta] \ ,
\end{equation}
where $A_\alpha$ is $N\times N$ Hermitian matrix corresponding to non-abelian gauge field. Finally, $P[E_{\alpha\beta}]$ is a pull-back of the background $E_{MN}(\Phi)$ defined as
\begin{equation}
P[E]_{\alpha\beta}=E_{\alpha\beta}+D_\alpha\Phi^r E_{r\beta}+E_{\alpha r}D_\beta\Phi^r+D_\alpha\Phi^r E_{rs}D_\beta\Phi^s  \ , 
\end{equation}
where $D_\alpha \Phi^m$ is covariant derivative
\begin{equation}
D_\alpha\Phi^m=\partial_\alpha \Phi^m+i[A_\alpha,\Phi^m] \ . 
\end{equation}
Note that non-abelian action for $N$ Dp-branes is implicitly defined in the static gauge where world-volume coordinates $\xi^\alpha$ coincide with the target space ones $x^\alpha$. 
Finally $\str$ means symmetrized trace and in order to describe infinite
number of Dp-branes we should divide the action (\ref{actnon}) by the infinite order the quotient
group $\mathbf{Z}^d$.

Further, $Q^i_{ \ j}$ is defined as
\begin{equation}
Q^i_{ \ j}=\delta^i_{ \ j}+i\lambda^{-1}[\Phi^i,\Phi_j]
\end{equation}
and $(Q^{-1})^j_{ \ k}$ is its inverse in the sense
\begin{equation}
Q^i_{ \ j}(Q^{-1})^j_{ \ k}=\delta^i_k \ . 
\end{equation}
Finally, $P[E_{\alpha r}E^{rs}((Q^{-1})_s^t-\delta_s^t)E_{t\beta}]$ is defined as
\begin{eqnarray}
& &P[E_{\alpha r}E^{rs}((Q^{-1})_s^t-\delta_s^t)E_{t\beta}]=\nonumber \\
& &E_{\alpha r}E^{rs}((Q^{-1})_s^t-\delta_s^t)E_{t\beta}+D_\alpha\Phi^m
E_{m r}E^{rs}((Q^{-1})_s^t-\delta_s^t)E_{t\beta}+\nonumber \\
& &E_{\alpha r}E^{rs}((Q^{-1})_s^t-\delta_s^t)E_{tk}D_\beta\Phi^k+
D_\alpha\Phi^m E_{m r}E^{rs}((Q^{-1})_s^t-\delta_s^t)E_{tn}D_\beta\Phi^n \ , 
\nonumber \\
\end{eqnarray}
where $E^{rs}$ is matrix inverse to $E_{mr}$ defined as
\begin{equation}
E_{mr}E^{rs}=\delta_m^s \ . 
\end{equation}
In order to implement T-duality along $d$ transverse directions we follow analysis performed in \cite{Brace:1998ku} which we generalize to the case of non-linear non-abelian action (\ref{actnon}). Let us presume that the background fields do not depend on  $x^A$ coordinates, where $A=p+1,\dots,p+d$ and that these coordinates are periodic with period $\sqrt{2\pi\lambda}$. 
This is natural if we recognize that all geometrical properties of the background are
encoded in the the field $E_{MN}$. Then we consider 
an infinite number of Dp-branes on compact space with coordinates $x^A$ when we impose 
following quotient conditions
\begin{eqnarray}\label{quoteq}
& &\mU_B^{-1}\Phi^A \mU_B=\delta^A_B\sqrt{\lambda}+\Phi^A \ , 
\nonumber \\
& &\mU_B^{-1}\Phi^{i'}\mU_B=\Phi^{i'} \ , \quad i'=p+d+1,\dots,9 \ .  \nonumber \\
\end{eqnarray}
Let us presume  that  solution of the 
quotient equation corresponds to operators $\mU_A$ that commute
\begin{equation}
[\mU_A,\mU_B]=0 \ . 
\end{equation}
In order to solve (\ref{quoteq}) it is natural to introduce 
an auxiliary Hilbert space on which $\Phi^A$ and $\mU_B$ act. The simplest way is to introduce Hilbert space of auxiliary functions living on $d-$dimensional torus taking value in $\mathbf{C}^d$. 
Then we take $\mU_A$ as generators of the functions on $d-$dimensional torus 
\begin{equation}
\mU_A=e^{i\lambda^{-1/2}\sigma_A}
\end{equation}
where $\sigma_A$ are coordinates on the covering space of torus. Then $\Phi^A$ has to be equal to
\begin{equation}\label{PhiA}
\Phi^A=-i\lambda(\partial^A-iA^A(\sigma_A))
\end{equation}
since then 
\begin{equation}
\mU_{B}^{-1}\Phi^A\mU_B=\lambda^{1/2}\delta_B^A+\Phi^A \ . 
\end{equation}
Using these results we can now proceed to write corresponding action in T-dual background. 
As the first step we perform following manipulation with the determinant in the action (\ref{actnon})
\begin{eqnarray}\label{detP}
& &\det (P[E_{\alpha\beta}]+P[E_{\alpha r}E^{rs}((Q^{-1})_s^t-\delta_s^t)E_{t\beta}]+
\lambda F_{\alpha\beta})\det Q^i_j=
\nonumber \\
& &=\det \left(\begin{array}{cc}
P[E]_{\alpha\beta}-P[E_{\alpha r}E^{rs}E_{s\beta}]+\lambda F_{\alpha\beta} & \bA_{\alpha}^{ \ n}\\
\bB^m_{\ \beta} &  Q^{mn}\\ \end{array}\right)\det E_{mn} \ , \nonumber \\
\end{eqnarray}
where $Q^{ij}=E^{ij}+i\lambda^{-1}[\Phi^i,\Phi^j]$ and where 
\begin{eqnarray}
\bA_\alpha^{ \ n}=D_\alpha\Phi^k E_{kr}E^{rn}+E_{\alpha r}E^{rn} \ , \quad 
\bB^m_{\ \beta}=-E^{mk}E_{k\beta}-D_\beta \Phi^m \ . 
\nonumber \\
\end{eqnarray}
First of all we observe that
\begin{eqnarray}
P[E_{\alpha\beta}-E_{\alpha r}E^{rs}E_{s\beta}]=
E_{\alpha\beta}-E_{\alpha r}E^{rs}E_{s\beta} \ . 
\nonumber \\
\end{eqnarray}
To proceed further  we use the fact that $D_\beta \Phi^A$ acting on arbitrary function  $f(\sigma)$ defined on the space labelled by $\sigma_A$ is equal to 
\begin{eqnarray}
D_\alpha\Phi^Af=\lambda(\partial_\alpha \Phi^A+i[A_\alpha,\Phi^A])f=\lambda(\partial_\alpha A^A-\partial^A A_\alpha)f\equiv \lambda F_\alpha^{ \ A}f \ 
\nonumber \\
\end{eqnarray}
and hence we can identify $D_\alpha \Phi^A$ with $\lambda F_\alpha^{ \ A}$. Using this identification we obtain 
\begin{eqnarray}
& &\bA_\alpha^{ \ A}=\lambda F_\alpha^{ \ A}+E_{\alpha r}E^{rA} \ , \quad 
\bA_\alpha^{ \ i'}=D_\alpha\Phi^{i'}+E_{\alpha r}E^{r i'} \ . 
\nonumber \\
& &\bB^A_{\ \beta}=-E^{AB}E_{B\beta}-E^{Ai'}E_{i'\beta}-\lambda F_\beta^{ \ A} \ ,
\nonumber \\
& &\bB^{i'}_{ \ \beta}=-E^{i'j'}E_{j'\beta}-E^{i' A}E_{A\beta}-\partial_\beta\Phi^{i'} \ \nonumber \\
\end{eqnarray}
and finally 
\begin{eqnarray}
& &Q^{AB}=E^{AB}+i\lambda^{-1}[\Phi^A,\Phi^B]=E^{AB}+\lambda F^{AB} \ , \nonumber \\
& &Q^{Ai'}=E^{Ai}+\lambda \partial^A\Phi^{i'} \ , \quad 
Q^{i'B}=E^{i'B}-\partial^B \Phi^{i'} \ , 
\quad Q^{i'j'}=E^{i'j'} \ , \nonumber \\
\end{eqnarray}
where we used (\ref{PhiA}) so that 
\begin{eqnarray}
[\Phi^A,\Phi^B]=-i\lambda^2 F^{AB}\ , 
[\Phi^A,\Phi^{i'}]=-i\lambda\partial^A\Phi^{i'} 
 \ . 
 \nonumber \\
 \end{eqnarray}
 Now we return to the first determinant in (\ref{detP}) and rewrite it to the form
 \begin{eqnarray}
& & \det \left(\begin{array}{cc}
 P[E]_{\alpha\beta}-P[E_{\alpha r}E^{rs}E_{s\beta}]+\lambda F_{\alpha\beta} & \bA_{\alpha}^{ \ n}\\
 \bB^m_{\ \beta} &  Q^{mn}\\ \end{array}\right)=\nonumber \\ 
 & &\det  \left(\begin{array}{ccc}
 E_{\alpha\beta}-E_{\alpha
 	 r}E^{rs}E_{st}+\lambda F_{\alpha\beta} & \bA_\alpha^{ \ B} & \bA_\alpha^{ \ j'} \\ 
 \bB^A_{\ \beta} & Q^{AB} & Q^{A j'} \\
 \bB^{i'}_{\ \beta} & Q^{i' B} & Q^{i' j'} \\ \end{array}\right)=
 \nonumber \\
 & &=\det  \left(\begin{array}{ccc}
E_{\alpha\beta}-E_{\alpha r}E^{rs}E_{s\beta}+\lambda F_{\alpha\beta} & \bA_\alpha^{ \ B} & \bA_\alpha^{ \ j'} \\ 
\bB^{A}_{ \ \beta}-Q^{Ak'}(Q^{-1})_{k' i'}\bB^{i'}_{\ \beta} & Q^{AB}-Q^{A k'}
(Q^{-1})_{k' i'}Q^{i'B} & 0 \\
\bB^{i'}_{\ \beta} & Q^{i' B} & Q^{i' j'} \\ \end{array}\right)=
\nonumber \\
& &=
\det  \left(\begin{array}{ccc}
E_{\alpha\beta}-E_{\alpha r}E^{rs}E_{s\beta}-\bA_\alpha^{ \ i'}
(Q^{-1})_{i'j'}\bB^{j'}_{ \ \beta}+\lambda F_{\alpha\beta} & \bA_\alpha^{ \ B}-\bA_{\alpha}^{ \ j'}
(Q^{-1})_{i'j'}Q^{j'B} & 0 \\ 
\bB^{A}_{ \ \beta}-Q^{Ak'}(Q^{-1})_{k' i'}\bB^{i'}_{\ \beta} & Q^{AB}-Q^{A k'}
(Q^{-1})_{k' i'}Q^{i'B} & 0 \\
\bB^{i'}_{\ \beta} & Q^{i' B} & Q^{i' j'} \\ \end{array}\right)\equiv
\nonumber \\
& &\equiv\det \left(\begin{array}{ccc} 
\bD_{\alpha\beta} & \bD_\alpha^{ \ B} & 0 \\
\bD^A_{ \ \beta} & \bD^{AB} & 0 \\
\bB^{i'}_{ \ B} & Q^{i'B} & Q^{i'j'} \\ \end{array}
\right)
  \end{eqnarray}
Since $Q^{i'j'}=E^{i'j'}$ it is clear that the matrix inverse $(Q^{-1})_{i'j'}$ is equal to
$(Q^{-1})_{i'j'}=\tE_{i'j'}$ where
\begin{equation}
\tE_{i'k'}E^{k'l'}=\delta_{i'}^{l'} \ . 
\end{equation}
Now we explicitly calculate components of the matrix $\bD$ as 
\begin{eqnarray}
& &\bD_{\alpha\beta}
=E_{\alpha\beta}-E_{\alpha r}E^{rs}E_{s\beta}
+\partial_\alpha \Phi^{ \ i'}\tE_{i'j'}E^{j'r}E_{r\beta}+
\partial_\alpha\Phi^{i'}\tE_{i'j'}\partial_\beta\Phi^{j'}+
\nonumber \\
& &+
E_{\alpha r}E^{r i'}\tE_{i'j'}E^{j'k}E_{k\beta}+
E_{\alpha r}E^{r i'}\tE_{i'j'
}\partial_\beta \Phi^{j'}+\lambda F_{\alpha\beta}=\nonumber \\
& &=E_{\alpha\beta}-E_{\alpha A}
(E^{AB}-E^{Ai'}\tE_{i'j'}E^{j'B})
E_{B\beta}+
\nonumber \\
& &+\partial_\alpha \Phi^{ \ i'}\tE_{i'j'}E^{j'r}E_{r\beta}+
\partial_\alpha\Phi^{i'}\tE_{i'j'}\partial_\beta\Phi^{j'}+
\nonumber \\
& &+
E_{\alpha r}E^{r i'}\tE_{i'j'
}\partial_\beta \Phi^{j'}+\lambda F_{\alpha\beta} \ . 
\nonumber \\	
\end{eqnarray}
To proceed further we observe that
\begin{equation}
(E^{AB}-E^{Ai'}\tE_{i'j'}E^{j'B})E_{BC}=\delta^A_C
\end{equation}
and hence we can identify expression in the bracket with the matrix
inverse $\tE^{AB}$ to $E_{AB}$ so that $\tE^{AB}E_{BC}=\delta^A_C$.
Further, let us consider following expression
\begin{equation}
E_{i'j'}-E_{i'A}\tE^{AB}E_{B j'}
\end{equation}
and multiply it with $E^{j'k'}$. Then, after some calculations, we get 
\begin{eqnarray}
(E_{i'j'}-E_{i'A}\tE^{AB}E_{B j'})E^{j' k'}=
\delta_{i'}^{k'} \nonumber \\
\end{eqnarray}
so that we can  identify expression in the bracket with  
matrix $\tE_{i'j'}$
\begin{equation}
\tE_{i'j'}=E_{i'j'}-E_{i'A}\tE^{AB}E_{B j'} \ . 
\end{equation}
Using these results we obtain following useful expressions
\begin{equation}\label{EAhelp}
E^{Ai'}\tE_{i'j'}=-\tE^{AB}E_{Bj'} \ , \quad  \tE_{i'j'}E^{j'B}=-E_{i'C}\tE^{CB}
\end{equation}
and
\begin{equation}\label{Ibetahelp}
\tE_{i'j'}E^{j'r}E_{r\beta}=
E_{i'\beta}-E_{i'A}\tE^{AB}E_{B\beta} \ . 
\end{equation}
Using (\ref{EAhelp}) and (\ref{Ibetahelp}) we get
\begin{eqnarray}
& &\bD_{\alpha\beta}=E_{\alpha\beta}-E_{\alpha A}\tE^{AB}E_{B\beta}+
\partial_\alpha\Phi^{i'}
(E_{i'\beta}-E_{i'A}\tE^{AB}E_{B\beta})
+(E_{\alpha}-E_{\alpha A}\tE^{AB}E_{B j'})\partial_\beta\Phi^{j'}+
\nonumber \\
& &\partial_\alpha\Phi^{i'}(E_{i'j'}-E_{i'A}\tE^{AB}E_{B j'})\partial_\beta
\Phi^{j'} \ \nonumber \\
\end{eqnarray}
and 
\begin{eqnarray}
& &\bD^{AB}=
\tE^{AB}+\lambda F^{AB}
+\partial^A\Phi^{i'}(E_{i'j'}-E_{i'A}\tE^{AB}E_{Bj'})\partial^B\Phi^{j'}-
\nonumber \\
& &-\tE^{AC}E_{Cj'}\partial^B \Phi^{j'}+\partial^A \Phi^{i'}E_{i'C}
\tE^{CB} \ . \nonumber \\
\end{eqnarray}
In the same way we obtain 
\begin{eqnarray}
& &\bD_\alpha^{ \ B}
=\lambda F_\alpha^{ \ B}+E_{\alpha A}\tE^{AB}+
\partial_{\alpha}\Phi^{i'}E_{i'C}\tE^{CB}+
\nonumber \\
& &+\partial_\alpha \Phi^{i'}(E_{i'j'}-E_{i'A}\tE^{AB}E_{Bj'})\partial^B\Phi^{j'}
+(E_{\alpha i'}-E_{\alpha C}\tE^{CD}E_{Di'})\partial^B\Phi^{i'}
\nonumber \\
\end{eqnarray}
and also
\begin{eqnarray}
& &\bD^A_{ \ \beta}=
-\tE^{AB}E_{B\beta}-\lambda F_\beta^{ \ B}-\tE^{AB}E_{Bj'}\partial_\beta\Phi^{j'}+
\nonumber \\
& &+\partial^A\Phi^{i'}(E_{i'\beta}-E_{i'B}\tE^{BC}E_{B\beta})+
\partial^A\Phi^{i'}\tE_{i'j'}\partial_\beta\Phi^{j'} \ .  \nonumber \\
\nonumber \\
\end{eqnarray}
Finally we consider following combinations of determinants that appear under square root in the action (\ref{actnon})
\begin{eqnarray}
& &\det E_{mn}\det E^{i'j'}=
\det (E_{i'j'}-E_{i'A}\tE^{AB}E_{Bj'})\det E_{AB} \det E^{i'j'}=
\nonumber \\
& &=\det \tE_{i'j'}\det E^{i'j'}\det E_{AB}=\det E_{AB} \ .
\nonumber \\ 
\end{eqnarray}
Collecting these terms together we obtain final form of D(p+d)-brane action in T-dual background in the form
\begin{equation}
S=-\frac{T_p}{\lambda^{d/2}}\int d^{p+1}\xi d^d\sigma
e^{-\tphi}\sqrt{-\det \bD} \ , 
\end{equation}
where we also used the relation between trace over infinite dimensional
matrices and integration over coordinates $\sigma$
\begin{equation}
\tr=\frac{1}{\lambda^{d/2}}\int d^d\sigma \  , 
\end{equation}
and where the matrix $\bD$ has following components
\begin{eqnarray}
& &\bD_{\alpha\beta}=\tE_{\alpha\beta}+\partial_\alpha \Phi^{i'}\tE_{i\beta}+
\tE_{\alpha j'}\partial_\beta\Phi^{j'}+\partial_\alpha\Phi^{i'}\tE_{i'j'}
\partial_\beta\Phi^{j'}+\lambda F_{\alpha\beta} \ ,  \nonumber \\
& &\bD_{\alpha}^{ \ B}=\tE_\alpha^{ \ B}+\partial_\alpha\Phi^{i'}\tE_{i'}^{ \ B}+
\tE_{\alpha j'}\partial^B\Phi^{j'}+\partial_\alpha\Phi^{i'}\tE_{i'j'}\partial^B\Phi^{j'}+\lambda F_\alpha^{ \ B} \ , \nonumber \\
& &\bD^A_{ \ \beta}=\tE^A_{ \ \beta}+\tE^A_{ \ j'}\partial_\beta\Phi^{j'}
+\partial^A\Phi^{i'}\tE_{i'\beta}+\partial^A\Phi^{i'}\tE_{i'j'}\partial_\beta\Phi^{j'} 
-\lambda F_\beta^{ \ A} \ , \nonumber \\
& &\bD^{AB}=\tE^{AB}+\partial^A\Phi^{i'}\tE_{i'j'}\partial^B\Phi^{j'}+\tE^A_{ \ j'}\partial^B \Phi^{j'}+\partial^A\Phi^{i'}\tE_{i'}^{ \ B}+\lambda F^{AB} \ , \nonumber \\
\end{eqnarray}
where we have following components of the background matrix $\tE$
\begin{eqnarray}\label{bacTdual}
& &\tE_{\alpha\beta}=E_{\alpha\beta}-E_{\alpha A}\tE^{AB}E_{B\beta} \ , \quad 
\tE_{i'\beta}=E_{i'\beta}-E_{i'A}\tE^{AB}E_{B\beta} \ , \nonumber \\
& &\tE_{\alpha j'}=E_{\alpha j'}-E_{\alpha A}\tE^{AB}E_{Bj'} \ , \quad 
\tE_{i'j'}=E_{i'j'}-E_{i'A}\tE^{AB}E_{B j'} \ , \nonumber \\
& &\tE_{\alpha}^{ \ B}=E_{\alpha A}\tE^{AB} \ ,  \quad 
\tE_{i'}^{ \ B}=E_{i'C}\tE^{CB} \ , \quad 
\tE_{i'j'}=E_{i'j'}-E_{i'A}\tE^{AB}E_{B j'} \ , \nonumber \\
& &\tE_{\alpha i'}=E_{\alpha i'}-E_{\alpha C}\tE^{CD}E_{Di'} \ , \quad 
\tE^A_{ \ \beta}=-\tE^{AB}E_{B\beta} \ ,  \nonumber \\
& &\tE_{i'j'}=E_{i'j'}-E_{i'A}\tE^{AB}\tE_{B j'} \ , \quad 
\tE^A_{ \ j'}=-\tE^{AC}E_{Cj'} \ , \quad \tE_{i'}^{ \ B}=
E_{i'C}\tE^{CB} \ , \nonumber \\
\end{eqnarray}
and where 
\begin{equation}
e^{-\tphi}=e^{-\phi}\sqrt{-\det E_{AB}}
\end{equation}
It is important to stress that resulting D(p+d)-brane propagates in T-dual background
where the T-dual background is given by Buscher's rules. More explicitly, note that we perform T-duality along directions labelled by $A,B,\dots$ where $A=p+1,\dots,p+d$ that should be identify with $m,n,\dots$ given in the section (\ref{second}). In the same way $\alpha,\beta=0,\dots,p$ and $i',j'=p+d+1,\dots,9$ should be identified with $\mu,\nu$ again given in section (\ref{second}). Then it is easy to see that (\ref{bacTdual}) coincide exactly with the T-duality rules given in (\ref{Tduallon}) and hence fully proves covariance of Dp-brane action under T-duality transformations. 

{\bf Acknowledgement:}
\\
This work 
is supported by the grant “Integrable Deformations”
(GA20-04800S) from the Czech Science Foundation
(GACR).

\newpage

\end{document}